# Metal-Ferroelectric-Metal structures with Schottky contacts: II. Analysis of the experimental current-voltage and capacitance-voltage characteristics of Pb(Zr,Ti)O$_3$ thin films

L. Pintilie[a], I. Boerasu
NIMP, P.O. Box MG-7, 76900 Bucharest-Magurele, Romania

M. J. M. Gomes
Dept. of Physics, University do Minho, Campus de Gualtar, 4710-057, Braga, Portugal

T. Zhao
Materials Research Science and Engineering Center, Univ. of Maryland, College Park, MD 20742, USA

R. Ramesh
Dept. of Materials Science and Engineering, Univ. of California at Berkeley, 210 Hearst Memorial Mining Building, Berkeley, CA 94720-1760

M. Alexe
Max Planck Institute for Microstructure Physics, Weinberg 2, 06120, Halle, Germany

*Abstract*
A modified model of metal-semiconductor contacts is applied to analyze the capacitance-voltage and current-voltage characteristics of metal-ferroelectric-metal structures. The ferroelectric polarization is considered as a sheet of surface charge situated at a fixed distance from the interface. The presumable high concentration of structural defects acting as active electric traps is taken into account by introducing a deep acceptor-like level. The model is applied on a set of metal-Pb(Zr,Ti)O$_3$-metal samples with different Zr/Ti ratios, deposited by different methods, and having different thickness, electrode materials, and electrode areas. Values around $10^{18}$ cm$^{-3}$ were estimated for the hole concentration from capacitance-voltage measurements. The space charge density near the electrode, estimated from current-voltage measurements, is in the $10^{20}$ - $10^{21}$ cm$^{-3}$ range. The total thickness of the interface layer ranges from 3 nm to 35 nm, depending on the Zr/Ti ratio, on the shape of the hysteresis loop, and on the electrode material. The simulated I-V characteristics is fitted to the experimental one using the potential barrier and Richardson's constant as parameters. The potential barrier is determined to be in the 1.09-1.37 eV range and the Richardson's constant is 520 Acm$^{-2}$K$^{-2}$.

---

[a] E-mail address: pintilie@mpi-halle.mpg.de

## Introduction

Electrical characterization of ferroelectric thin films used in microelectronic applications such as memory devices, usually comprises current-voltage (I-V) and capacitance-voltage (C-V) characteristics. Information regarding charge transport mechanisms and charge distribution within the ferroelectric film can in principle be extracted from these measurements.[1] Several models were proposed to interpret and to simulate experimental I-V curves. Most of these models assume the presence of a potential barrier at the metal-ferroelectric interface, similar to the case of a Schottky contact, but the ferroelectric film is regarded as completely depleted. This is the particular case of the Pb(Zr,Ti)O$_3$ (PZT) films with Pt contacts, which is one of the most studied systems.[2-4] A separate interpretation of I-V and C-V characteristics was possible, but the results of the two type of measurements were hardly ever correlated.[5-14]

A model of the metal-ferroelectric interface, based on well-known theories for metal-semiconductor Schottky contacts, was recently developed. Unlike the previous models of metal-ferroelectric-metal structures, this model considers the effect of the ferroelectric polarization on the energy bands bending near electrode interfaces, as well as the contribution of the electrically active traps to the electric field value in the interface region.[15]

The aim of the present paper is to present a methodology to extract the relevant quantities of a Schottky contact, in the specific case of metal-ferroelectric-metal structures, by using the above mentioned model. The model has been applied to PZT films with different Zr/Ti ratio, fabricated by different deposition methods, and having different Schottky contacts. In all cases the experimental I-V and C-V characteristics are discussed and the relevant parameters of the structure were estimated from the analysis of the experimental results.

## Model summary

The recently proposed model[15] is based on similar models applied for metal-semiconductor Schottky contacts.[16-19] The ferroelectric material is considered as a wide-gap p-type semiconductor with the ferroelectric polarization modeled as a sheet of superficial charge located at a fixed finite distance from the metal-ferroelectric interface. A deep acceptor-like level is also considered. A schematic of the structure introducing the main quantities is shown in Fig. 1

The characteristic quantities of the Schottky contact were then deduced by solving Poisson's equation in a similar way with that used in case of a p-type CdTe epitaxial layer with surface charge at the interface with the substrate.[18,19] The equation giving the apparent built-in potential $V_{bi}'$, the depletion layer width $w$, the depletion layer specific capacitance $C$, and the maximum electric field at the interface $E_m$ are given by:[15]

$$V_{bi}' = V_{bi} \pm \frac{P}{\varepsilon_0 \varepsilon_{st}} \delta \quad (1)$$

$$V_{bi} = \Phi_B^0 - \frac{kT}{q} Ln\left(\frac{N_V}{p(T)}\right) \quad (2)$$

$$w = \sqrt{\frac{2\varepsilon_0 \varepsilon_{st}(V + V_{bi}')}{qN_{eff}}} \quad (3)$$

$$C = \frac{\varepsilon_0 \varepsilon_{st}}{w} = \sqrt{\frac{q\varepsilon_0 \varepsilon_{st} N_{eff}}{2(V + V_{bi}')}} \quad (4)$$

$$E_m = \sqrt{\frac{2qN_{eff}(V + V_{bi}')}{\varepsilon_0 \varepsilon_{st}}} \pm \frac{P}{\varepsilon_0 \varepsilon_{st}} \quad (5)$$

The quantities involved in the above equations are: $V_{bi}$ – normal built-in potential, in the absence of the ferroelectric polarization; $P$ - ferroelectric polarization; $\delta$ - thickness of the interface layer; $\varepsilon_0$ - permitivitty of the free space; $\varepsilon_{st}$ - low frequency (static) dielectric constant of the ferroelectric layer; $q$ - electron charge; $k$ - Boltzmann's constant; $T$ - temperature; $N_V$ - effective density of states in the valence band; $p(T)$ - free hole concentration at the temperature $T$; $\Phi_B^0$ - potential barrier at zero voltage; and $N_{eff}$ - effective charge density in the depleted region taking into consideration all the charges, including the charged traps.

The above equations were then used to interpret the current-voltage (I-V) and capacitance voltage (C-V) characteristics and to develop a methodology to extract the main parameters of the structure: $P$, $V_{bi}'$, $N_{eff}$, $\delta$, $p(T)$, and $\Phi_B^0$. It was argued that the thermionic emission over the potential barrier is the most probable injection mechanism at room temperature.[15] Therefore, the equation of the current density is given by:[16]

$$J = A^* T^2 \exp\left(-\frac{q}{kT}\left(\Phi_B^0 - \sqrt{\frac{qE_m}{4\pi\varepsilon_0\varepsilon_{op}}}\right)\right) \quad (6)$$

where $A^*$ is Richardson's constant, and $\varepsilon_{op}$ is the high frequency dielectric constant. For sufficiently large applied voltages the linear term in polarization in (5) can be neglected and the Schottky representation becomes $\ln(J) \sim (V+V_{bi}')^{1/4}$. The Schottky representation is a straight line with the slop $b$ given by:

$$b = \frac{q}{kT}\sqrt[4]{\frac{q^3 N_{eff}}{8\pi^2\varepsilon_0^3\varepsilon_{op}^2\varepsilon_{st}}} \quad (7)$$

If the dielectric constants are known, the effective charge density in the depleted region $N_{eff}$ can be estimated from (7). The apparent built-in potential $V_{bi}'$ is chosen to obtain the best linear fit on the considered voltage range.

From the C-V characteristics the "doping" density can be evaluated, according to:[16,20]

$$N_{dop} = \frac{2}{q\varepsilon_0\varepsilon_{st}\left[d(1/C^2)/dV\right]} \quad (8)$$

The above affirmation is valid only if all the doping impurities are ionized at room temperature and only if the concentration of deep levels is negligible compared with that of the doping impurities. At this point it has to be underlined that what is evaluated from the C-V measurements is not a fixed charge density as it is usually assumed but a mobile charge density.[20] The density calculated using (8) is in fact the density of the free carriers from the neutral volume able to follow the small ac probing voltage used to measure the capacitance. In the case of a p-type semiconductor $N_{dop} = p(T)$. A clear distinction has to be made between the theoretical equation of the depletion layer capacitance (4), in which a fixed charge is involved ($N_{eff}$) and the principle of the capacitance measurements in which the ac probing field can be followed only by the mobile charges. Therefore, one might expect that $N_{eff} \neq p(T)$ if there is a large contribution of deep traps to the effective charge density. Depending on their emission and capture time constants the traps might not follow the high frequency ac field used in the capacitance measurement.

## Methodology to extract the relevant quantities of a metal-PZT-metal structure with Schottky contacts.

Basically, three measurements are necessary to characterize a metal-ferroelectric-metal structure with Schottky contacts at a certain temperature. These are the polarization – electric field (P-E) hysteresis loop, the current-voltage (I-V), and the capacitance-voltage (C-V) characteristics. The I-V measurements at different temperatures can add information regarding the conduction mechanism.

The hysteresis loop gives the polarization and coercive field values as well as the static dielectric constant Additionally, it shows for a given ferroelectric how far from ideality the ferroelectric switching is. A rectangular hysteresis loop is understood to represent the ideal ferroelectric switching. It has to keep in mind that the real measured quantity is the electric displacement $D$ and not directly the ferroelectric polarization. The linear part $\varepsilon_0\varepsilon_{st,l}E$ of the electric displacement $D$ has to be extracted from the measured quantity in order to obtain the real field dependence of the ferroelectric polarization. Therefore, the knowledge of the low frequency linear dielectric constant $\varepsilon_{st}$ is necessary. The correct evaluation of the linear part of the dielectric constant is from the saturated part of the hysteresis loop, when the ferroelectric polarization is fully switched and only the linear polarization contributes to a further increase of the electric displacement $D$. However, it is not always possible to obtain saturation due to sample breakdown. The high losses or the high leakage can also make of difficult to identify the linear part of the hysteresis loop. In these unfavorable cases, the dielectric constant calculated from the capacitance measured at zero bias on fresh samples and with very small ac amplitude is an acceptable approximation.

The I-V characteristics give information on the effective charge density $N_{eff}$ and the apparent built-in voltage $V_{bi}'$. From the Schottky representation $\ln(J) \sim (V+V_{bi}')^{1/4}$ the effective $N_{eff}$ can be estimated using (7). As $V_{bi}'$ cannot be estimated from independent measurements it should be considered as a fit parameter.[15] Therefore, the $V_{bi}'$ value should be adjusted to obtain the best linear fit for the considered voltage range. This value can be subject to relatively large errors, up to 50 %, but it can nevertheless give an idea of the polarization influence on the normal built-in potential.

From the C-V characteristics the free hole concentration $p(T)$ can be estimated using (8). $p(T)$ can also be roughly estimated from the slope of the $1/C^2 \sim V$ representation, if the doping is uniform.

The critical field $E_m^{cr}$ at the interface is given by:[16]

$$E_m^{cr} = \frac{A^* T^2}{q N_V \mu} \tag{9}$$

If $E_m$ given by (5) is higher than $E_m^{cr}$ then the current is of purely thermionic nature, i.e. controlled only by the barrier, and equation (6) applies.

The experimental I-V curve in case of pure thermionic emission can be simulated using the determined values for $V_{bi}'$ and $N_{eff}$ and keeping as parameters the potential barrier height at zero voltage $\Phi_B^0$ and Richardson's constant $A^*$. Since it enters into the exponential term in (6), the potential barrier height $\Phi_B^0$ has the largest effect on the shape of the I-V curve, as well as on the current magnitude. Richardson's constant $A^*$ slightly influences the magnitude of the current. Knowing $\Phi_B^0$ and $p(T)$, the normal built-in potential $V_{bi}$ can be estimated using (2) and finally, using $V_{bi}$, $V_{bi}'$ and $P$ in (1) the interface layer thickness $\delta$ can be estimated.

In summary, the methodology to extract the basic parameters of the metal-ferroelectric structures with Schottky contacts has the following steps:
- Hysteresis loop measurement. The polarization value is evaluated.
- I-V measurements. From the $\ln(J) \sim (V+V_{bi}')^{1/4}$ representation $V_{bi}'$ and $N_{eff}$ are estimated. The knowledge of both static and dynamic dielectric constants is necessary in this case. The static dielectric constant can be evaluated from direct capacitance measurements at zero bias, on fresh samples. The dynamic dielectric constant can be evaluated from optical measurements.
- With $N_{eff}$, $V_{bi}'$, $P$ and the dielectric constants the maximum electric field at the interface $E_m$ is calculated and the I-V characteristic is simulated using the potential barrier $\Phi_B^0$ and Richardson's constant as fit parameters. The hole effective mass can be calculated from $A^* = A_0^*(m^*/m_0)$, where $A_0^*$ is Richardson's constant for the free electron, $m^*$ is the effective mass, $m_0$ is the free-electron mass.
- C-V measurements. The free carrier (holes) density $p(T)$ is estimated
- The normal built-in potential $V_{bi}$ is calculated using (2).
- From $V_{bi}'$, $V_{bi}$ and $P$ the thickness of the interface layer $\delta$ is estimated.

In the followings this methodology will be applied to real metal-ferroelectric structures. The experimental I-V and C-V characteristics will be analyzed and discussed in light of the proposed model.

## *Experimental*

A number of lead zirconate-titanate PZT films have been analyzed using the above model and methodology. The samples have different Zr/Ti ratios i.e. from 92/8 to 20/80, thickness from 150 nm to 400 nm, and different electrode materials such as Pt and $SrRuO_3$. The samples were prepared by different methods such as sol-gel (SG), metalorganic decomposition (MOD), and pulsed-laser deposition (PLD). Sample details are summarized in Table 1.

In all cases the perovskite structure was confirmed by X-ray diffraction (not shown here). The samples deposited by SG and MOD are polycrystalline.[21,22] The sample PZT20/80 is a high-quality epitaxial as reported elswhere.[23,24]

The characterization methods employed for this study were:
- Polarization-electric field (P-E) hysteresis loop acquired at 1 kHz using an AixACCT TF Analyzer 2000 .
- Current-voltage (I-V) measurements were performed using electrometers Keithley 617 and Keitley 6517. The applied voltage was stepwise swept from zero to the desired value and, in order to reach a steady state, a delay time of 10 up to 30 seconds was used setting the voltage step and the current reading. Before any I-V measurement the films were pre-poled by applying for 1 minute a dc field higher than the coercive field. The poling voltage has always the same polarity as the voltage used in the I-V measurements. In this way the polarization orientation, as well as its value are set and the contribution of the current due to polarization reversal is minimized. This contribution is negligible in the case of materials with rectangular hysteresis loops.
- Capacitance-voltage (C-V) measurements were performed at 1 kHz using either a HP4263A LCR bridge, or HP 4194A impedance analyzer.

All measurements were performed at room temperature. In some cases the dynamic dielectric constant was determined from optical transmission-reflectance measurements performed on films deposited under the same conditions on MgO substrates. The measurements were performed using a UV-VIS-NIR spectrometer and the obtained value is about 6.5.[25]

## *Results*

As was mentioned in the above-presented methodology, hysteresis measurements were first performed and the results are shown in Figs. 2 and 3. As expected, the investigated films have very different ferroelectric properties. The shape of the loops is different from sample to sample. The polarization values are scattered over almost a decade. Typically, ferroelectric properties of epitaxial films are better than those of polycrystalline films. This can be also observed in our epitaxial sample, which exhibits a square-like hysteresis loop. The SG and MOD samples show slim loops with a relatively large asymmetry. Polarization is obviously larger for epitaxial films than for the polycrystalline films. The saturation polarization, which is used in the theoretical model, along with the other evaluated quantities are given in Table 2. The hysteresis loop closest to an ideal rectangular loop is shown by the PZT20/80 sample. The sample PZT92/8 is on the opposite side, being far away from ideality. However, the model has been applied even in this unfavorable case when the leakage current has most probably contributions from the polarization reversal. The fact that the polarization is not constant during the I-V measurement, being dependent on the applied voltage, makes the apparent built-in potential, the thickness of the interface layer and the effective charge in the depleted region also field-dependent, due to divP ≠ 0. Nevertheless, the results obtained using a constant polarization are helpful to get a general picture on the analyzed samples and to make a comparison between the structures that are close or far from ideality.

The free carrier (holes) concentration as function of the applied bias can be evaluated from the capacitance-voltage (C-V) characteristics by plotting the derivative of $1/C^2$ vs. $V$ according to (8). The experimental C-V characteristics are presented in Figs. 4, 5 and $d(1/C^2)/dV$ vs. $V$ plots are shown in Figs. 6 and 7 for the extreme cases PZT92/8 and PZT20/80, the two samples that are the closest and the furthest from ideality, respectively regarding the hysteresis loop. The plot shows the doping profile across the film thickness as in the case of normal semiconductors.[20] As can be seen from Figs. 6 and 7, except the region around zero applied bias, the hole concentration estimated using equation (8) is constant with the applied voltage, indicating a constant doping throughout the film thickness. The voltage range for which (8) can be applied is much smaller in the case of sample PZT92/8, due to the fact that the polarization reversal is taking place on a larger voltage range in this case. As expected, the influence of the polarization reversal on the measured capacitance is much higher in such cases in which the hysteresis loop is far from ideality. Only when the polarization is relatively saturated equation (8) is valid and allows good estimates of the hole concentration. The estimated values are given in the Table 2 for all the samples. In all cases the estimated hole concentration is in the $10^{18}$ cm$^{-3}$ range. The opposite signs for the hole concentration on the two voltage polarities are due to the change in the sign of the $d(1/C^2)/dV$ derivative.

Current-voltage (I-V) characteristics and the $\ln(J) \sim (V+V_{bi}'')^{1/4}$ Schottky plots are presented in Fig. 8 and 9, respectively. In all cases the I-V characteristics are not symmetric although the top and bottom electrodes are identical. This is most probably due to different thermal histories of the two electrodes leading to different properties of the Schottky contacts. The Schottky representation is linear in all cases and for both polarities. Only the voltage range over which this representation is valid differs from sample to sample.

The effective charge density in the depleted region near the electrodes was evaluated using (7). As was mentioned in the previous section, the apparent built-in potential is chosen in order to obtain the best linear fit on the maximum voltage range. The obtained values for $V_{bi}'$ and $N_{eff}$ are also presented in Table 2 along with the static dielectric constant determined from capacitance measurements at 1 kHz. The values obtained for $N_{eff}$ were not identical for the two interfaces suggesting a non-uniform charge distribution. However, in the following a constant $N_{eff}$ will be considered, and the value will be taken as the average value at the two interfaces. It has to be noted that negative values of $V_{bi}'$ were obtained in some cases. According to equation (1), negative values for $V_{bi}'$ are possible, depending on the polarization value and on the distance $\delta$ between the polarization charge and the physical interface with the electrode. These odd results were obtained on polycrystalline samples containing extended structural defects, such as grain boundaries, that might affect the interface properties. Hence, these should be regarded with caution. Nevertheless, the presence of the polarization charge drastically reduces the built-in potential at the reverse-biased contact in all cases including the epitaxial film.

With the above determined main parameters one can fit the I-V characteristics using (6), and considering the potential barrier height $\Phi_B^0$ and Richardson's constant $A^*$ as fitting parameters. Although the fit procedure was performed for all the samples, only the case of PZT20/80 will be discussed in detail. The result of the fit procedure is shown in Fig. 10, which presents the experimental and the calculated I-V characteristics. The maximum electric field at the interface used in (6) was calculated using (5), with $V_{bi}'$, $N_{eff}$, and $P$ determined from the experimental I-V and P-E curves. The values obtained for the potential barrier are given in Table 2. The potential barrier and Richardson's constant for the PZT20/80 films were found to be about 1.30 eV and 520 Acm$^{-2}$K$^{-2}$, respectively. The same value of $A^*$ was then used to simulate the I-V characteristics of all samples, keeping only the potential barrier as fitting parameter. The values for all film compositions are also given in Table 2.

The effective density of states in the valance band $N_V$ used in equation (2) to compute the normal built-in potential was then estimated using the effective mass obtained from the above given Richardson's constant. The resulting value of $4.3m_0$, where $m_0$ is the mass of the free electron, is close to the value used in other models for materials with perovskite structure.[6,7]

## *Discussions*

Despite its simplicity and many approximations, the proposed model for the metal-ferroelectric interface with Schottky contacts describes successfully the I-V characteristics and, in combination with C-V measurements, allows evaluation of the relevant parameters of the metal-ferroelectric Schottky structure as well as of the ferroelectric film. Nevertheless, there are some limitations in applying the model. One of these limitations regards the depletion layer width. The model is valid only if at low voltages the total width of the depletion layers is less than half of the film thickness in order to properly calculate the maximum electric field at the interface using (5) and then the current density using (6).[16,17] Otherwise, the film is quickly depleted and the electric field within the film becomes constant. The voltage dependence of the depletion width $w$, calculated using (3), is shown in Fig. 11 for the extreme Zr/Ti ratios. As MFM structures have always two depletion layers at zero bias, when an external bias is applied the depletion layer width near the reverse-biased contact increases, whereas the depletion layer width near the forward-biased contact will decrease. Additionally, due the presence of the ferroelectric polarization the two depletion widths are not equal at zero bias because the apparent built-in potentials are not the same for the two interfaces, according to equation (1). For the investigated samples the total width of the depleted regions is lower than half the film thickness even for the sample PZT92/8, which has the lowest effective charge density in the depleted region. According to equation (3), the lower $N_{eff}$, the higher the depletion width will be. We mention occasionally that the depletion layer width has to be calculated considering the effective charge density in the depleted layer and not the charge density deduced from C-V measurements, which in fact is a density of free carriers.

An additional condition that has to be fulfilled regards the thermionic emission and the validity of equation (6). This equation (6) can be used only if the maximum electric field at the interface is higher than a certain critical field calculated using (9). If the maximum field $E_m$ given by (5) is higher than the critical field then the emission is purely termionic and the current is controlled by the barrier height. If $E_m$ is lower than the critical field, the injection takes place by thermionic emission but the current is limited by the bulk properties. The last case might be possible in insulating materials with low free carrier concentration, or in materials with very low carrier mobility. In the present work the estimation of the critical field was done using a mobility value of 400 cm$^2$/Vs, which was recently reported.[26] The effective density of states in the valance band was calculated using the above-estimated value for the effective mass. The value obtained for the critical field is about 4 kV/cm. For comparison, the lowest value for the maximum field at the interface of the reverse-biased contact, calculated at zero applied voltage using equation (5) and using the values from Table 2, is about 500 kV/cm. Comparing with the estimated value for the $E_m^{cr}$ it can be seen that the condition $E_m > E_m^{cr}$ is always fulfilled. Thus the equation (6) holds for the investigated voltage range even the mobility is lower than the considered value.

The Scottky representation $ln(J) \sim (V+V_{bi}'')^{1/4}$ and equation (7) were deduced and used in the above analysis assuming that the constant $P/\varepsilon_0\varepsilon_{st}$ term in equation (5) is negligible compared with the first term for the whole voltage range. This is a crude approximation and normally introduces errors in the estimation of the effective charge density $N_{eff}$, especially at low voltages. Neglecting the constant term in (5) introduces an error in the estimation of the maximum electric field of about 30 % at low voltages. The errors drops below 10 % at the highest voltage used in the I-V measurements. These errors in $E_m$ can give errors of about 40 % in the estimation of $N_{eff}$ using (7). Even with these relatively high errors the order of magnitude given in Table 2 remains valid. The estimation of the potential barrier $\Phi_B^0$ will also be affected by an error of about 0.1 eV when the polarization term in (5) is neglected. Nevertheless, these errors do not affect the main conclusions drawn on the basis of the model.

Regarding the experimental I-V curves presented in Fig. 9, it can be seen that in case of PZT92/8 and PZT65/35 the current increases suddenly at a certain voltage for both polarities. It is not the purpose of this paper to explain the I-V characteristics in every detail, but it can be assumed that a change in the conduction mechanism occurs at these voltages. Possible causes can be an increased probability for tunneling through the tip of the potential barrier or breakdown of the barrier. In the first case the field emission becomes predominant, whereas in the second case space charge limited currents become important. The initiation of time dependent breakdown might also be possible.

The increase of the effective charge density in the depletion layer with increasing the Zr/Ti ratio is a surprising result. It would be expected for the epitaxial PZT20/80 to have the lowest density of structural defects acting as traps, whereas the sol-gel deposited or MOD samples to have a higher density of defects. However, it has to be noticed that both the hole concentration $p(T)$ and effective charge concentration $N_{eff}$ in the depletion layer increases with increasing the Zr/Ti ratio. This fact suggests that the Zr-rich samples are more resistive. On the other hand, the acceptor-like level assumed in the model to explain the discrepancy between the I-V and C-V results has a higher density in Ti-rich samples. A relation between the Ti ions and the acceptor-like level can be assumed. It can be speculated that this trapping level is related with $Ti^{4+} \rightarrow Ti^{3+}$ transition, or is associated with a complex involving Ti which charges negatively by capturing an electron.[27] The activation energy for Ti-related level was reported to be about 0.7 eV.[28] This level is active at room temperature, having an activation energy comparable with that of defect levels known to be active at room temperatures in materials with about the same band gap as PZT.[29,30]

It is worth noting that the quality of the ferroelectric film is an important factor. The interface layer thickness seems to be dependent on the hole concentration and on the saturation polarization value. It appears that these factors are interrelated: the better the quality of the film, the higher the hole concentration. Consequently, the polarization is higher, the internal polarization screening is better, and the interface layer is thinner. However, this observation needs to be confirmed by further correlated studies of structure and electrical properties for films of different qualities but with the same Zr/Ti ratio.

The estimated potential barriers are in good agreement with those already reported for Pt/PZT and $SrRuO_3$/PZT interfaces.[1,14,25,31,32] The thickness of the interface layer is of the order of magnitude assumed for the "dead-layer", at least for the PLD-deposited sample[33-35]. This fact supports the idea that the better the crystalline quality of the film (epitaxial layer) the thinner is the interface layer. It is interesting to observe that the evaluated thickness for PZT20/80, of about 3 nm, is very close to the critical thickness estimated theoretically for ferroelectric thin films.[36] Below the critical thickness the ferroelectric polarization can no longer be detected or does not exist any more.

Regarding the values for $N_{eff}$, it might appear odd that the trap concentration is higher in the epitaxial film than in the polycrystalline ones. Nevertheless, we should pay attention to the kind of defects existing in the two type of films. In epitaxial PZT only point defects are expected, except some misfit dislocations at the interface with the bottom electrode. Such point defects will have a well defined energetic position in the band gap and will act as traps for a certain type of carriers. A polycrystalline film usually has a large amount of extended defects, such as grain boundaries or compositional non-homogeneities. In this case the energetic spectrum of the associated trap levels can extend over the entire forbidden band. These levels can capture both electrons and holes, changing the effective charge in the SCR and leading to a completely different behaviour than in the case of epitaxial films. It has to be emphasized that a high polarization is associated with a high value of $N_{eff}$. It can be assumed that the trapped charges might play an important role in the compensation of the polarization charge.

## *Conclusions*

The model proposed for the metal-ferroelectric Schottky interface and the methodology developed for the evaluation of relevant structure quantities from I-V and C-V measurements was applied to PZT films with different Zr/Ti ratios. The model considers the presence of the ferroelectric polarization as a sheet of surface charge at a finite distance from the electrode interface. An acceptor-like level was additionally introduced to account for the presumed high density of intrinsic defects acting as traps. Using the same methodology for all films the apparent built-in potential $V_{bi}'$, the effective charge density in the depletion region $N_{eff}$, and the hole concentration $p(T)$ were evaluated from the experimental I-V and C-V curves. By fitting the theoretical I-V curve with the experimental data it was possible to obtain the potential barrier height at zero bias and Richardson's constant. The following values were obtained: $2\times10^{17}$ cm$^{-3}$ - $5\times10^{18}$ cm$^{-3}$ for the hole density; $3.4\times10^{19}$ cm$^{-3}$ – $1.7\times10^{21}$ cm$^{-3}$ for the effective space charge $N_{eff}$; 3.0 – 33.0 nm for the thickness of the interface layer; 1.09 – 1.37 eV for the potential barrier at zero field.

The simulated I-V characteristics fit well the experimental data in case of the studied samples. The fact that the same formalism works well for ferroelectric films with different Zr/Ti ratio prepared at different locations, by different methods, different electrodes and sample geometry, is a strong argument in support of the present model, although the measurements were performed only at room temperature. It can be concluded that the proposed model is valid for PZT capacitors regardless of their composition and deposition procedure, providing that potential barriers exists at the metal-PZT interfaces.

## *Acknowledgements*
Part of this work was supported by Volkswagen Stiftung within the project 'Nano-sized Ferroelectric Hybrids' under contract No. I/77737 and I/80897 and partly by NATO under project SfP-971970 and grant CP(RO)04/C/2001/PO.
The authors are grateful to Prof. Ulrich Goesele for the useful discussion.

Table 1 Preparation details of the studied samples.

| PZT | Preparation | Thickness (nm) | Annealing | Electrodes |
|---|---|---|---|---|
| 92/8 | Sol-Gel[1] | 200 | CTA (650 $^0$C, 30 min) in oxygen | Pt electrodes, 0.5 mm$^2$ |
| 65/35 | Sol-Gel[1] | 150 | CTA (650 $^0$C, 30 min) in air | Pt electrodes, 0.5 mm$^2$ |
| 55/45 | Sol-Gel[1] | 150 | CTA (650 $^0$C, 30 min) in air | Pt electrodes, 0.5 mm$^2$ |
| 30/70 | MOD[2] | 400 | CTA (650 $^0$C, 60 min) in oxygen | Pt electrodes, 0.275 mm$^2$ |
| 20/80 | PLD[3] | 240 | | SrRuO$_3$, 8x10$^{-4}$ mm$^2$ |

[1] Univ. do Minho, Braga, Portugal
[2] MPI-Halle, Germany
[3] Univ. of Maryland, USA
CTA-conventional thermal annealing

Table 2. The evaluated values for saturation polarization $P_S$, static dielectric constant $\varepsilon_{st}$, hole concentration $p(T)$, apparent built-in potential $V_{bi}'$, effective space charge density in the depleted layer $N_{eff}$, potential barrier $\Phi_B^0$, and interface layer thickness $\delta$.

| PZT | $P_S$ (μC/cm$^2$) | $\varepsilon_{st}$ | $p(T)$ (10$^{17}$ cm$^{-3}$) | $V_{bi}'$ (V) | $N_{eff}$ (average) (10$^{20}$ cm$^{-3}$) | $\Phi_B^0$ (average) (eV) | $\delta$ (average) (nm) |
|---|---|---|---|---|---|---|---|
| 92/8 | 6.1 | 247 | 3.8 | +0.10 | 0.42 | 1.21 | 33.2 |
| 65/35 | 22.3 | 303 | 5.0 | +0.05 | 1.40 | 1.21 | 13.3 |
| 55/45 | 31.1 | 438 | 2.2 | -0.10 | 0.34 | 1.09 | 12.1 |
| 30/70 | 16.0 | 280 | 22 | -0.20 | 3.4 | 1.37 | 22.6 |
| 20/80 | 40.5 | 180 | 50 | +0.40 | 17 | 1.30 | 3.0 |

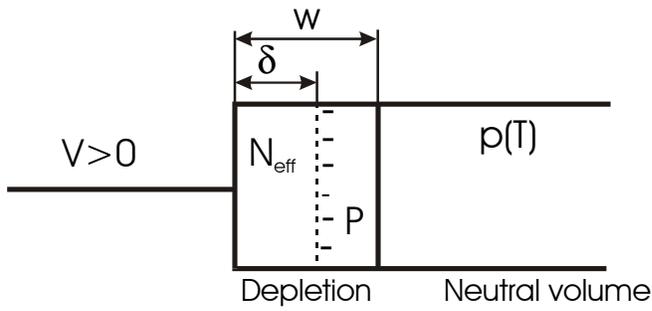

Fig. 1 Schematic of the metal-ferroelectric structure where $w$ is the depletion layer width, $\delta$ - the distance from the polarization sheet of charge to the physical metal-ferroelectric interface, $p(T)$ – the concentration of the free carriers in the neutral volume, $N_{eff}$ – the charge density in the depleted region, and $P$ – ferroelectric polarization.

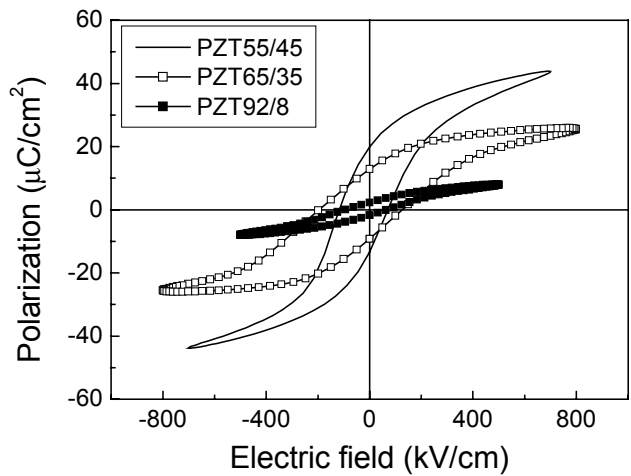

Fig. 2 Ferroelectric hysteresis loops of the sol-gel deposited PZT92/8, PZT65/35, and PZT55/45 films.

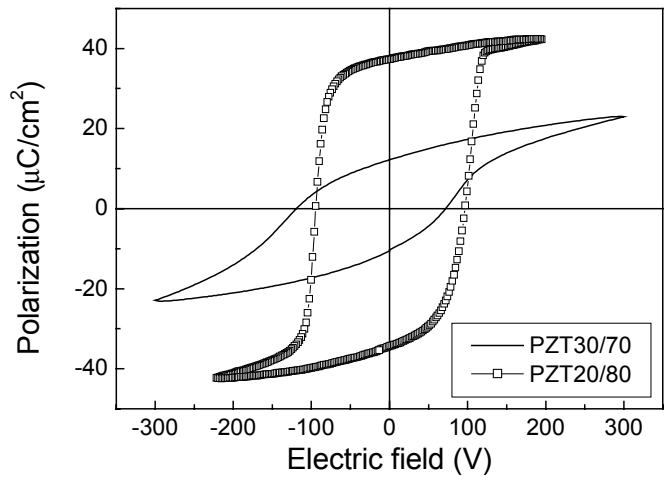

Fig. 3 Ferroelectric hysteresis loop of the MOD-prepared PZT30/70 film and for the PLD-deposited PZT20/80 film.

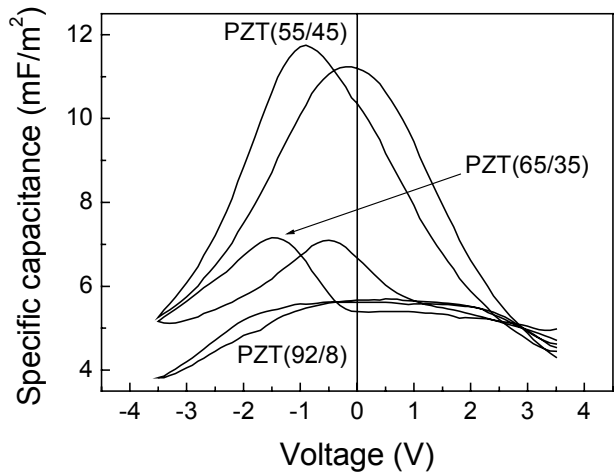

Fig. 4 Capacitance-voltage characteristics of the sol-gel-deposited PZT92/8, PZT65/35, and PZT55/45 films.

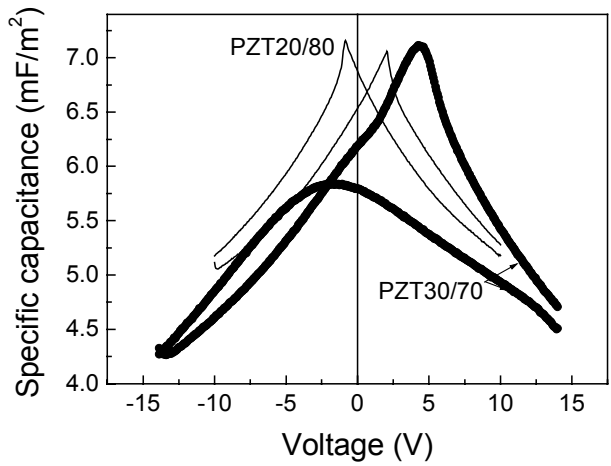

Fig. 5 Capacitance-voltage characteristics of the MOD-deposited PZT30/70 and the PLD-deposited PZT20/80 films, respectively.

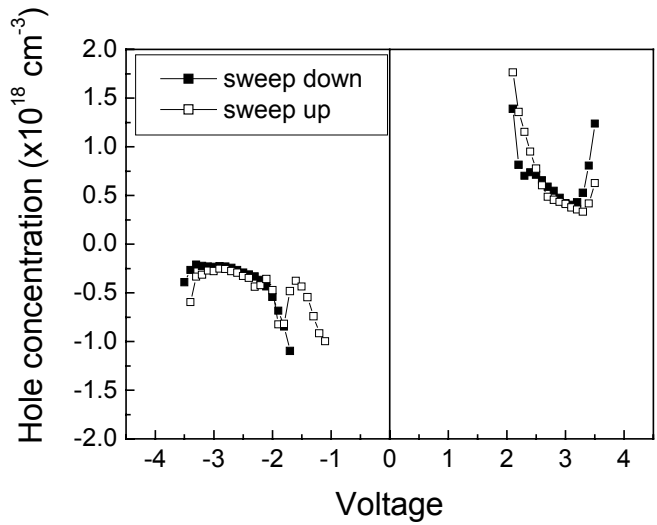

Fig. 6 Hole concentration estimated by applying equation (8) to the experimental C-V characteristics of the sample PZT92/8.

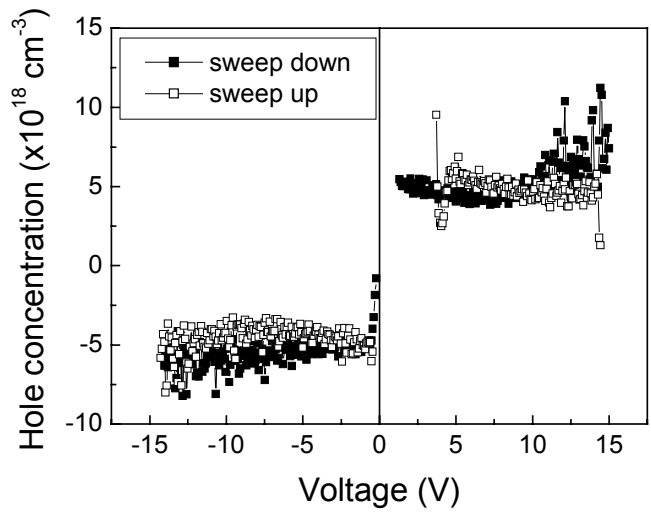

Fig. 7 Hole concentration estimated by applying equation (8) to the experimental C-V characteristics of the sample PZT20/80.

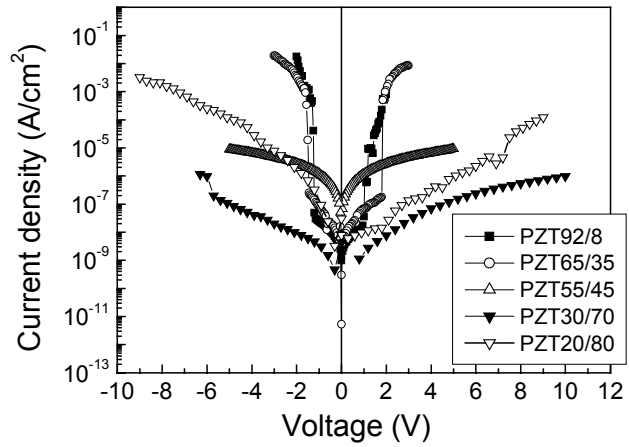

Fig. 8 Current-voltage characteristics of all studied films.

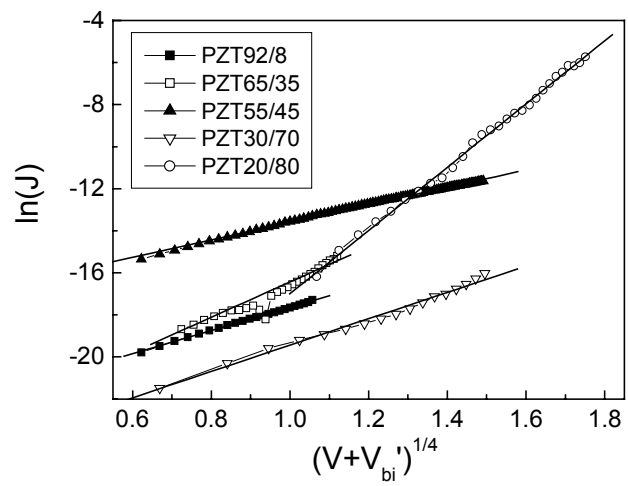

(a)

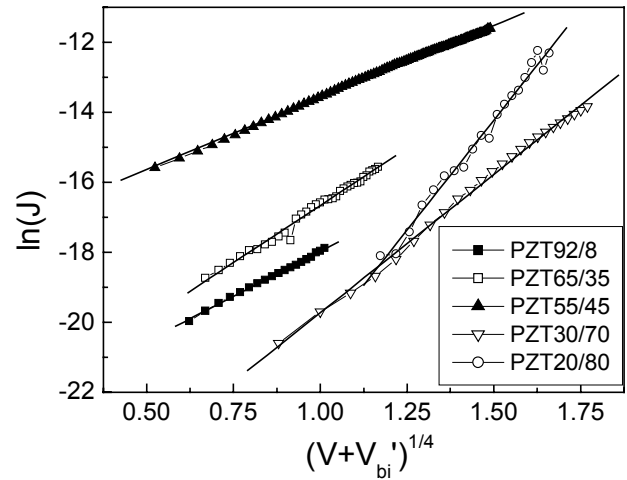

(b)

Fig. 9 Schottky representations for (a) negative polarity and (b), positive polarity of the applied bias.

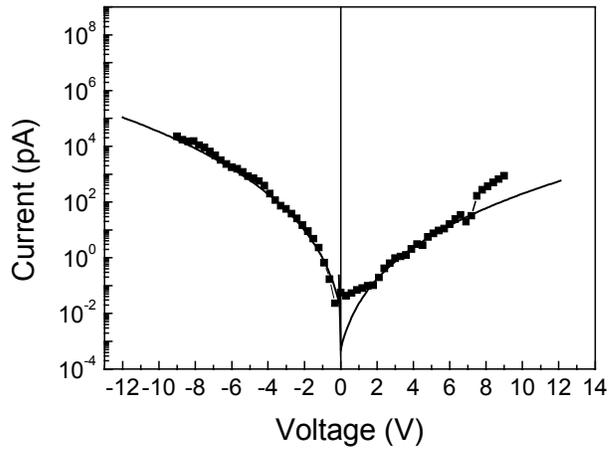

Fig. 10 Experimental data (■) and simulated I-V characteristic (—) for the epitaxial PZT20/80 film.

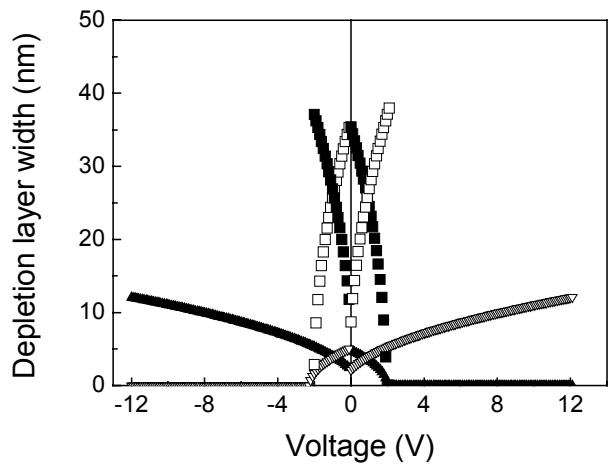

Fig. 11 Voltage dependence of the depletion width calculated using equation (3).